\documentclass[aps,prl,twocolumn]{revtex4}

\usepackage{natbib}
\usepackage{amssymb}
\usepackage{amsthm}
\usepackage{graphicx}
\usepackage{graphics}
\usepackage{subfigure}
\newcommand{\ket}[1]{|#1\rangle}

\def\beas{\begin{eqnarray*}}
\def\eeas{\end{eqnarray*}}
\def\bea{\begin{eqnarray}}
\def\eea{\end{eqnarray}}
\def\be{\begin{equation}}
\def\ee{\end{equation}}

\begin{document}

\title{Coulomb blockade as a probe for non-Abelian statistics in Read-Rezayi states}%

\author{Roni Ilan}
\affiliation{Department of Condensed Matter Physics, Weizmann
Institute of Science, Rehovot 76100, Israel}

\author{Eytan Grosfeld}
\affiliation{Department of Condensed Matter Physics, Weizmann
Institute of Science, Rehovot 76100, Israel}

\author{Ady Stern}
\affiliation{Department of Condensed Matter Physics, Weizmann
Institute of Science, Rehovot 76100, Israel}

\date{\today}

\begin{abstract}
We consider a quantum dot in the regime of the quantum Hall
effect, particularly in Laughlin states and non-Abelian
Read-Rezayi states. We find the location of the Coulomb blockade
peaks in the conductance as a function of the area of the dot and
the magnetic field. When the magnetic field is fixed and the area
of the dot is varied, the peaks are equally spaced for the
Laughlin states. In contrast, non-Abelian statistics is reflected
in modulations of the spacing which depend on the magnetic field.

\end{abstract}


\maketitle

% ----------------------------------------------------------------
%
\begin{figure}
\includegraphics[width=0.28\textwidth]{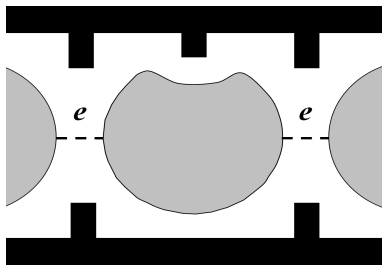}
\caption{Fabry Perot interferometer in the limit of strong
quasiparticle backscattering. The dot (a quantum Hall droplet) is
coupled to the leads via electron tunneling, and its area may be
varied using a side modulation gate. }\label{fig:inter}
\end{figure}

The possibility that quasiparticles in certain quantum Hall
systems satisfy non-Abelian statistics has been widely discussed
in the last two decades \cite{MR,ReadRezayi, TQCreview,
AnyonsReview}. This exciting theoretical possibility, however, has
at present no experimental support (in fact, it is only very
recently that first steps towards experimental tests of {\it
Abelian} fractional statistics have been carried out
\cite{camino}). Furthermore, only a few predictions have been put
forward that may experimentally identify non-Abelian
quasiparticles. The current paper contributes to bridging this gap
between theory and experiment by predicting signatures of
non-Abelian statistics on Coulomb blockade peaks in transport
through quantum dots.

Most experiments proposed so far for observing non-Abelian
statistics rely on interference of quasiparticle trajectories
\cite{Fradkin,sarma,SternHalperin,kirill1,kirill2,Stone, feldman,
feldman2} in a two point contact geometry that creates Fabry-Perot
or Mach-Zehnder interferometers. The Fabry-Perot interferometer
\cite{chamon}, sketched in Fig.\ref{fig:inter}, is a Hall bar with
two quantum point contacts (QPCs) introducing back-scattering
through quasiparticle tunnelling from one edge to the other. In
lowest order interference experiments
\cite{SternHalperin,kirill1,kirill2,Stone}, the sensitivity to the
statistics of quasiparticles originates from the motion of the
back-scattered current around quasiparticles that are localized in
the bulk, in the ``island" formed between two interfering
trajectories. The number of these quasiparticles, $n_{is}$, may be
varied in a controlled way. For fixed $n_{is}$, signatures of
non-Abelian statistics manifest themselves in the interference
term of the backscattered current. When the number $n_{is}$
fluctuates in time, signatures of non-Abelian statistics may be
present in the current noise \cite{Eytan}.

In this work we study the limit of strong back-scattering by the two
point contacts (see Fig.\ref{fig:inter}). In this limit, the area
between the point contacts becomes a quantum dot, weakly coupled to
the rest of the Hall bar. The low temperature conductance through
the dot is suppressed by its charging energy, except in the
degeneracy points that give rise to Coulomb blockade peaks
\cite{GellerLoss}. We show that for non-Abelian quantum Hall states
of the Read-Rezayi series \cite{ReadRezayi}, the position of the
peaks in a two-parameter plane of the area of the dot, $S$, and the
magnetic field, $B$, is sensitive to the non-Abelian statistics of
the quasiparticles. Here the origin of the sensitivity is the effect
of the localized quasiparticles on the spectrum of edge excitations.
Such a sensitivity was already discovered for the $\nu=5/2$ state in
\cite{SternHalperin}, and we will compare our results for the
Read-Rezayi states to those of the $\nu=5/2$ state.

The Read-Rezayi states are expected to occur in the filling factor
range of $2<\nu<3$. We assume that the point contacts strongly
back-scatter only the edge state of the uppermost, partially
filled Landau level. The quantum dot is defined then by the edge
state of the partially filled Landau level. At the end of the
paper we discuss the case in which all edge states are
back-scattered by the two point contacts.

For an almost closed quantum dot the number of electrons is quantized
to an integer, and the low-voltage low-temperature conductance through
the dot is suppressed unless the ground state energy of the dot with $N$
electrons is degenerate with its ground state energy with $N+1$ electrons.
Thus, Coulomb blockade peaks of the conductance appear for those values
of the area and magnetic field for which the following equation
\begin{equation}
E(N,S,B)=E(N+1,S,B) \label{coulomb-peaks}
\end{equation}
is satisfied for some integer $N$.

For a clean large ($N\gg 1$) dot in a metallic state at zero
magnetic field, where the electronic density is determined by
charge neutrality with a uniform positive back-ground of density
$n_0$, one expects the area that separates consecutive Coulomb
blockade peaks to be $\Delta S=e/n_0$, the area occupied by one
electron. This would also be the situation for the quantum Hall
state of non-interacting electrons at $\nu=1$. Below, we start by
showing that this is also the expected spacing for the Abelian
Laughlin $\nu=1/p$ states (with $p$ odd). In contrast, for the
Read-Rezayi series we find a much richer structure, that depends
on $B$: while the \emph{average} spacing between peaks remains
$e/n_0$, the presence of non-Abelian quasiparticles in the bulk
imposes modulations of the spacing which depend on their number,
$n_{is}$. The latter is determined by the magnetic field.

For all quantum Hall states the bulk is incompressible, and
electronic transport takes place along the edge. The energies in
(\ref{coulomb-peaks}) are then energies of edge modes. The edge of
the Abelian Laughlin states is described \cite{wen} by the action
of a chiral free boson (we take $\hbar=1$)
\begin{equation}
S=-\frac{1}{4\pi\nu}\int dt dx \left [
\partial_t\varphi\partial_x\varphi+v_c(\partial_x\varphi)^2\right ]
\label{action-abelian}
\end{equation}
where $v_c$ is the velocity of edge excitations. The bosonic field
$\varphi(x)$ is normalized here such that the electronic creation
operator is $e^{ip\varphi(x)}$, and the electron density along the
edge is given by $\rho=\frac{1}{2\pi}\partial_x \varphi$. For the
electron operator to be single valued the field $\varphi$ must
obey the quantization condition
\begin{equation}\label{eq:quantization}
\varphi(L)-\varphi(0)=2\pi n/p.
\end{equation}
The total number of electrons on the edge is $N=n\nu=n/p$.
Alternatively, the number of quasiparticles is $n$. Since the number
of electrons in the dot is an integer, the total number of
quasiparticles in the dot (edge and bulk together), which is
$n+n_{is}$, must be divisible by $p$.

When the magnetic field $B=B_{0}$ is such that the filling fraction
is precisely $1/p$ there are no quasiparticles in the bulk. The
energy is then $E=\frac{v_c}{4\pi\nu}\int dx(\partial_x\varphi)^2$,
which is minimized by a space-independent $\varphi(x)$. As the area
of the dot is varied continuously, the field $\varphi$ is restricted
by the quantization condition (\ref{eq:quantization}), and therefore
may change only in a discrete way. Therefore, an infinitesimal
increase in the area of the dot violates charge neutrality at the
edge, with an associated energy cost. When the area grows
sufficiently, it becomes energetically favorable to add a whole
electron to the dot. The energy dependence on the area may then be
incorporated into the description (\ref{action-abelian}) by writing
the edge energy as
\begin{equation}
E_c=\frac{v_c}{4\pi\nu}\int
dx\left(\partial_x\varphi-2\pi\nu\frac{B_{0}(
S-S_0)}{L\phi_0}\right)^2,
\end{equation}
where $\phi_0$ is the magnetic flux quantum. The total energy is
minimized when the charge density is uniformly spread along the
perimeter of the edge, i.e., when $\partial_x\varphi$ is
$x$-independent. With $N=0$ defined to be the number of electrons
for a dot with area $S_0$, the energy for $N$ electrons on the
edge is
\begin{equation}
E_c(N)=\frac{\pi v_c}{\nu L}\left(N-\nu\frac{B_{0}(
S-S_0)}{\phi_0}\right)^2. \label{ec}
\end{equation}
Eq. (\ref{coulomb-peaks}) reduces to $E_c(N)=E_c(N+1)$, and the area
separation $\Delta S$ between its solutions for consecutive values
of $N$ is $\Delta S=e/n_0$. This value is independent of the
magnetic field, as long as the bulk is incompressible: as the
magnetic field is changed from $B_{0}$ quasiparticles enter the
bulk. The incompressibility of the bulk quantizes their number to an
integer $n_{is}$. As a consequence, the number of quasiparticles on
the edge, $n$ of Eq. (\ref{eq:quantization}), is not necessarily an
integer multiple of $p$, but rather an integer of the form $\ell
p-n_{is}$, with $\ell$ an integer. That does not, however, change
the area spacing $\Delta S$. As we will now see, this is not the
case for non-Abelian states.

While the free chiral boson field
theory of Eq.(\ref{action-abelian}) fully describes the edge of a
$\nu=1/p$ state, the edge of the Read-Rezayi non-Abelian states
requires also a second field theory, whose properties we now
review. The second theory is a Conformal Field Theory (CFT) of a
neutral field, and for the $\nu=2+k/(k+2)$ Read-Rezayi state (with
$k=2,3,4...$), it is the $Z_k$ parafermionic CFT. Quasiparticles
for this state have charge $\frac{e}{k+2}$. When the magnetic
field is varied by one flux quantum, $k$ quasiparticles appear,
hence the flux associated with a single quasiparticle is
$\frac{2\pi}{ke}$.

The creation operators of both an electron and a quasiparticle are
then products of two factors. The first, $e^{i\alpha\varphi(z)}$,
accounts for the flux and the charge associated with the electron
($\alpha=(k+2)/k$), and with the quasiparticle ($\alpha=1/k$). The
second part is a neutral field labelled by two quantum numbers
$\Phi^l_m$, obeying the restrictions $l\in \left\{ 0,1,\ldots
,k\right\}$, $\Phi _{m }^{l }=\Phi _{m +2k}^{l}=\Phi _{m -k}^{k-l
}$ and $l+m \equiv 0\left( \text{mod}2\right) $. The integer $m$
is known as the holomorphic charge (or $Z_k$ charge) of the field
$\Phi _{m }^{l }$. Using the above identifications, the integer
$m$ may be restricted to the range $-l<m\leq l$. Fields that
deserve special mention are the identity, $I=\Phi _{0 }^{0 }$, the
parafermions $\psi_l=\Phi^0_{2l}$, and the parafermionic primary
fields, also known as spin fields, $\sigma_l\equiv \Phi^l_l$,
since the electron creation operator is $\psi_1
e^{i\frac{(k+2)}{k}{\varphi}}$ while the quasiparticle creation
operator is $\sigma_1 e^{i\frac{1}{k}{\varphi}}$. The fusion rules
for the parafermion CFT fields are given by
\cite{ZF,Gepner:1986hr}
\begin{equation}\label{eq:fusion}
\Phi _{m _{\alpha}}^{l _{\alpha}}\times \Phi _{m _{\beta}}^{l
_{\beta}}=\sum\limits_{l =\left| l_{\alpha}-l _{\beta}\right|
}^{\min \left\{l _{\alpha}+l_{\beta},2k-l _{\alpha}-l
_{\beta}\right\} }\Phi _{m _{\alpha}+m _{\beta}}^{l }.
\end{equation}

The conformal dimensions of the fields $\Phi _{m}^{l}$ , which
will be crucial in determining the energy spectrum, are
$h_{m}^{l}=\frac{l(l+2)}{4(k+2)}-\frac{m^2}{4k}$. The conformal
dimension of the bosonic sector is $\nu\alpha^2/2$. The
short-range product of two fields, known as the operator product
expansion (OPE), is given by
\begin{eqnarray}
\Phi _{m _{\alpha}}^{l _{\alpha}}(w)\times \Phi _{m _{\beta}}^{l
_{\beta}}(z)=\sum_{l_\gamma}C_{\alpha \beta \gamma}(z-w)^{\Delta
h}\ \Phi _{m _{\alpha}+m _{\beta}}^{l_\gamma }\label{ope}
\end{eqnarray}
where the fields appearing on the right hand side are determined
by equation (\ref{eq:fusion}), $C_{\alpha \beta \gamma}$'s are
constants, and $\Delta h=h_{m _{\alpha}+m _{\beta}}^{l_\gamma
}-h_{m _{\beta}}^{l _{\beta}}- h_{m _{\alpha}}^{l _{\alpha}}$. As
a consequence of that relation, when a field $\Phi _{m
_{\alpha}}^{l _{\alpha}}$ goes around a field $\Phi _{m
_{\beta}}^{l _{\beta}}$ and their fusion is to a field $\Phi _{m
_{\alpha}+m _{\beta}}^{l_\gamma }$, the phase associated is
$2\pi\Delta h$.

Let us now use this general input from the theory of CFT to
calculate the spectrum of the edge. First we consider the case when
the bulk of the dot does not include any quasiparticles
($n_{is}=0$). The fusion rules (\ref{eq:fusion}) imply that $k$
parafermions $\psi_1$ fuse to the identity field. This property of
the $Z_k$-theory captures the clustering of the electrons in the
Read-Rezayi states into groups of $k$-electrons \cite{ReadRezayi}.
We imagine starting with the total number of electrons in the dot
being divisible by $k$, and the system being relaxed into its ground
state. As the number of electrons is varied, the remainder, which
may assume any value between $1$ and $k-1$ electrons, accumulates at
the edge. The parafermionic state of the edge is then obtained by
applying $j$ operators $\psi_1$ to the vacuum, with $0\leq j\leq
k-1$. The energy of that state, denoted $E_\psi$, is calculated in
the following way.

The Hilbert space of parafermionic states is constructed by acting
with creation modes of the parafermion $\psi_1$ on the vacuum
\cite{Gepner:1986hr,mathieu2,Bouw}. Although the $1+1$ dimensional
geometry of the edge may be thought of as a cylinder described by
a single coordinate, $\xi=v_nt+ix$, where $v_n$ is the velocity of
the neutral sector as it propagates along the edge, it is easier
to work on the punctured plane by performing a conformal
transformation of the coordinates,
$z=e^{2\pi\xi/L}$\cite{CFTbook}. On the plane, the parafermion
$\psi_1$ is expanded in modes as follows
\begin{equation}\label{eq:mode expansion}
\psi_1=\sum_m z^{-m-h^0_{2}}\psi^1_{m},
\end{equation}
with $h_{2}^0=1-1/k$ being its conformal dimension. The allowed
values of the index $m$ are determined by the boundary conditions
imposed on $\psi_1$ by the field it acts on
\cite{ZF,Gepner:1986hr}. In this case, since it acts on the
vacuum, $\psi_1$ has periodic boundary conditions. Therefore, we
must have $m\in\mathbb{Z}+1/k$. However, if $\psi_1$ acts on an
edge that already contains a parafermion, as in
$\psi^1_{m_2}\psi^1_{m_1}\ket{vac}$, then when it encircles the
already existing parafermion, it accumulates also a phase of
$2\pi(h_{4}^0-2h_{2}^0)=-4\pi/k$. Then, the allowed values of
$m_2$ are $m_2\in\mathbb{Z}+3/k$. Similarly, for an edge that
contains $j$ parafermions, the allowed modes for the $j+1$
parafermion are $m_{j+1}\in\mathbb{Z}+(2j+1)/k$.

Since $[\mathcal{L}_0,\psi_m^1]=-m\psi_m^1$, where $\mathcal{L}_0$
is the Virasoro algebra generator proportional to the Hamiltonian
$H=\frac{2\pi v_n}{L} \mathcal{L}_0$, states created by repetitive
applications of $\psi^1$ modes on the vacuum are eigenstates of
the Hamiltonian. A general state with $j$ parafermions is of the
form
\begin{equation}\label{eq:state}
\psi^1_{-p_{j}+(2j-1)/k}\psi^1_{-p_{j-1}+(2j-3)/k}\cdot\cdot\cdot\psi^1_{-p_1+1/k}\ket{vac},
\end{equation}
where the $p$'s are integers. The eigenvalue of $\mathcal{L}_0$
for such a state is $\sum_{i=1}^j (p_i-(2i-1)/k)$. States with
negative eigenvalues have zero norm and are un-physical.

In Refs.~\cite{mathieu2,Bouw} it was shown that by choosing the
integers $p$ in Eq.(\ref{eq:state}) such that $p_{i+1} \geq p_i
\geq 1$, the set of states obtained is free of zero norm vectors.
Therefore the lowest energy state with $j$ parafermions is
obtained by choosing $p_i=1$ for all $i$. Under these constraints,
the lowest allowed value for the energy is therefore
\begin{equation}\label{eq:energies}
E_\psi(j)=\frac{2\pi v_n}{L}\frac{j(k-j)}{k}.
\end{equation}

To obtain the energy of the state with $j$ electrons on the edge,
we must sum $E_\psi$ and the contribution of the bosonic field,
$E_c$, given by equation ($\ref{ec}$) with $N=j$ and $\nu=k/(k+2)$
(the filling fraction of the uppermost partially filled Landau
level).

Given the expression $(\ref{eq:energies})$, together with Eqs.
(\ref{ec}) and (\ref{coulomb-peaks}), we can extract the area
spacing $\Delta S$,
\begin{equation}
\Delta S=\frac{e}{n_0}+\frac{eL \nu}{2 n_0 \pi
v_c}\left[E_\psi(N+2)-2E_\psi(N+1)+E_\psi(N)\right]. \label{area}
\end{equation}
The second term, which is central to our discussion,  adds a $k$
dependent modulation to the average spacing $e/n_0$, and will have
two possible values: since $j$ of equation (\ref{eq:energies}) is
restricted to be in the range $0\leq j\leq k-1$ while $N$ is not,
the spacing is given by
\begin{equation}\label{eq:spacing1nisZero}
\Delta
S_1=\frac{e}{n_0}\left(1-\nu\frac{v_n}{v_c}\frac{2}{k}\right)\end{equation}
as long as $[N+1]_k\neq 0$. When $[N+1]_k= 0$, the spacing is
larger and given by
\begin{equation}\label{eq:spacing2nisZero}
\Delta
S_2=\frac{e}{n_0}\left(1+\nu\frac{v_n}{v_c}\left(2-\frac{2}{k}\right)\right).\end{equation}
The pattern observed will be a bunching of the Coulomb blockade
peaks into groups of $k$ peaks. Within a group, the peaks are
separated by $\Delta S_1$, while the area spacing between
consecutive groups will be $\Delta S_2$. This $k$-periodicity of
the area spacing reflects the construction of the Read-Rezayi
states from clusters of $k$ electrons.

The effect of $n_{is}$ bulk quasiparticles on the bosonic part of
the edge theory is, similar to the Abelian case, a change in the
boundary conditions on $\varphi$. That change does not affect
$E_c(N+1)-E_c(N)$. The parafermionic energy $E_\psi$ depends on
$n_{is}$, since the presence of quasiparticles in the bulk changes
the boundary conditions for the field $\psi_1$ on the edge, and
hence its spectrum. We now analyze this effect in detail, and show
that it makes $\Delta S$ depend on $n_{is}$.

According to the fusion rules (\ref{eq:fusion}), $n_{is}$
quasiparticles in the bulk, each created by the operator $\sigma_1
e^{i\frac{1}{k}{\varphi}}$, will fuse to a combination of fields of
the form $\Phi_{\tilde{n}}^ae^{i\frac{n_{is}}{k}{\varphi}}$, where
${\tilde{n}}=[n_{is}]_k\equiv n_{is}(\text{mod}k)$ and the possible
values of $a$ are determined by (\ref{eq:fusion}). Since we start
with $N(\text{mod}k)=0$, the ground state has $a=\tilde{n}$.

When the parafermionic part of the bulk quasiparticles fuse to
$\Phi_{\tilde{n}}^{\tilde{n}}=\sigma_{\tilde{n}}$, the edge is not
in a vacuum state even when all electrons on the dot are clustered
to clusters of $k$. Rather, the state of the edge is
$\Phi_{k-\tilde{n}}^{k-\tilde{n}}\ket{vac}=\sigma_{k-\tilde{n}}\ket{vac}$.
The boundary conditions on a $\psi_1$ operating on this highest
weight state are then
\begin{equation}\psi_1(ze^{2\pi i})=e^{2\pi
i\frac{\tilde{n}-k}{k}}\psi_1(z).
\end{equation}
Accordingly, the modes $m$ in the expansion (\ref{eq:mode
expansion}) are $m\in\mathbb{Z}+(k+1-\tilde{n})/k$. Again, the
lowest energy state with a single $\psi_1$ mode is created by the
creation operator with the smallest value of $|m|$, with $m$
itself being non-positive. Similarly to the $n_{is}=0$ case, the
allowed values of $m$ change with the number of parafermions on
the edge, and for the $j$'th parafermion become
$m\in\mathbb{Z}+(2j-1+k-\tilde{n})/k$, where the value of $j$ is
limited by $k-1$. Due to the presence of a non-trivial $Z_k$
charge of the highest weight state
$\sigma_{k-\tilde{n}}\ket{vac}$, there will be another restriction
on the integers $p_1,...,p_j$ of Eq.(\ref{eq:state}): for $i>
\tilde n>0$ we must choose $p_{i}\geq 2$ \cite{mathieu2}.

Again, the energy $E_\psi$ for $j=[N]_k$ parafermions is
determined by the sum of the indices $m_i$ for $i=1..j$. This sum
depends on $\tilde n$ and therefore on $n_{is}$,
\begin{eqnarray}\label{eq:energiesnis}
E_\psi(j,\tilde n)=\frac{2\pi v_n}{L}h_\sigma+\frac{2\pi
v_n}{L}\times\left\{\begin{array}{cc}
                                    \frac{j(\tilde n-j)}{k} & j\leq \tilde n \\
                                    \frac{(j-k)(\tilde n-j)}{k} & j> \tilde n \\
                                  \end{array}\right.
\end{eqnarray}
Where $h_\sigma$ is the zero point energy of the spectrum,
determined by the conformal dimension of the relevant primary
field $\sigma_{k-\tilde{n}}$ acting on the vacuum.

Substituting Eq.  (\ref{eq:energiesnis}) in Eq. (\ref{area}) we
may study the spacings between Coulomb blockade peaks through the
properties of the spectrum. We again find that the peaks bunch
into groups, however, this time they do not bunch into groups of
$k$, but rather into alternating groups of $\tilde{n}$ and
$k-\tilde{n}$ peaks. The spacing that separates peaks within a
group is again given by Eq.(\ref{eq:spacing1nisZero}), while the
spacing that separates two consecutive groups is
\begin{equation}\label{eq:spacing2nisZero}
\Delta
S_2=\frac{e}{n_0}\left(1+\nu\frac{v_n}{v_c}\left(1-\frac{2}{k}\right)\right).\end{equation}
Therefore, for an odd value of $k$, the only possible period of
the peak structure is $k$, while when $k$ is even and
$[n_{is}]_k=k/2$ we find a periodicity of $k/2$.

For $k=2$ this result reproduces the even-odd effect predicted to
occur at $\nu=5/2$ in Ref.\cite{SternHalperin}: for odd $n_{is}$
the periodicity will be $k/2=1$, while for even $n_{is}$ it will
become $k=2$.

The reflection of non-Abelian statistics in the magnetic field
dependence of $\Delta S$ carries over to the case when the point
contacts back-scatter also the two edge states of the two filled
Landau levels. In that case the peaks we analyzed are
super-imposed on peaks associated with tunnelling to the edge
states of the filled levels. The spacing $\Delta S$ of the latter
does not depend on magnetic field, however, and may therefore be
separated from the ones of the partially filled level
\cite{SternHalperin}.

To summarize, we calculated the position of the Coulomb blockade
peaks on the two parameter plane of the area and magnetic field. For
a fixed value of the magnetic field, we found that for the Laughlin
states, the spacing between peaks is $\Delta S=e/n_0$. For
Read-Rezayi states the peaks form groups of $k$ peaks, where each
group splits into two subgroups, one containing $[n_{is}]_k$ peaks
and the other containing $k-[n_{is}]_k$ peaks. Having a period of
$k$ for $\Delta S$ is a consequence of the clustering of electrons,
similar to the case of a super-conductor, where the spacings between
Coulomb blockade peaks alternate between two values due to the
energy cost associated with having an unpaired electron. The
dependence on $n_{is}$ and the periodicity of $k/2$ occurring for
even $k$ and $[n_{is}]_k=k/2$ are unique aspects of the non-Abelian
nature of the quasiparticles.

We thank D. Gepner for instructive discussions. We acknowledge
support from the US-Israel Binational Science Foundation and the
Minerva foundation. \hspace{-0.5in}

%\bibliography{nonabelian}

\end{document}